%% file: paper.tex
\documentclass[12pt]{article}
\usepackage{amsfonts,amsbsy}
\newcommand{\be}{\begin{equation}}
\newcommand{\ee}{\end{equation}}
\newcommand{\bea}{\begin{eqnarray}}
\newcommand{\eea}{\end{eqnarray}}

\newcommand{\no}{\\ \nonumber}
\newcommand{\tAtwo}{\tilde{\mathbf{A}}_2}
\newcommand{\bM}{\mathbf{M}}
\newcommand{\bO}{\mathbf{O}}
\newcommand{\cS}{{\cal S}}
\newcommand{\cC}{{\cal C}}
\newcommand{\cM}{{\cal M}}
\newcommand{\Sh}{\widehat{S}}
\newcommand{\rmga}{(r-\gamma_{\alpha})}
\newcommand{\rpga}{(r+\gamma_{\alpha})}
\newcommand{\NC}{N_c}
\newcommand{\NS}{N}
\newcommand{\SUN}{SU(N_c)}
\newcommand{\CTT}{\mathbf{\widetilde{C}_2}}
\newcommand{\CTE}{\mathbf{\widetilde{C}_3}}
\newcommand{\VM}{{\cal A}}

\newcommand{\GV}[1]{{\cal V}(#1)}

\input{epsf.tex}

\begin{document}

\vskip -4cm

\begin{flushright}
FTUAM-98-24

IFT-UAM/CSIC-99-4
\end{flushright}


{\Large
\centerline{{\bf $\NS=1$ Supersymmetric Yang-Mills on the lattice    }}
\centerline{{\bf   at strong coupling}}
\vskip 0.2cm
}
{\large
\centerline{ 
E.~Gabrielli{$^{\dag \ddag}$}
\footnote{{\em e-mail:} egabriel@daniel.ft.uam.es},  
A. Gonz\'alez-Arroyo{$^{\dag \ddag}$} 
\footnote{{\em e-mail:} tony@martin.ft.uam.es} and  
C.~Pena{$^{\dag}$}
\footnote{{\em e-mail:} carlos@martin.ft.uam.es}
}
}
\vskip 0.3cm

\centerline{{$^\dag$}Departamento de F\'{\i}sica Te\'orica C-XI}
\centerline{Universidad Aut\'onoma de Madrid,}
\centerline{Cantoblanco, Madrid 28049, SPAIN.}
\vskip 10pt
\centerline{{$^\ddag$}Instituto de F\'{\i}sica Te\'orica C-XVI,}
\centerline{Universidad Aut\'onoma de Madrid,}
\centerline{Cantoblanco, Madrid 28049, SPAIN.}

\vskip 0.6cm

\begin{center}
{\bf ABSTRACT}
\end{center}
We study $\NS=1$  supersymmetric $\SUN$  Yang-Mills theory on the lattice
at   strong coupling and large $\NC$. Our method is based on the hopping 
parameter expansion in terms of random walks, resummed  for any value 
of the Wilson parameter $r$ in the  small  
hopping parameter region. Results are given for the mesonic (2-gluino)
and fermionic (3-gluino) propagators and spectrum. 
\vskip 0.8 cm
\begin{flushleft}

\end{flushleft}

\newpage

\input{section1.tex}

\input{section2.tex}

\input{section3.tex}

\input{section4.tex}

\input{acknow.tex}

\input{apendiceA}

\newpage

\begin{table}

\label{tb:tabla1}

\begin{center}
\caption{Nomenclature and  normalisation for the spin $\cS_i$ and colour $\cC_i$ 
matrices used to define the four dimensional 2-gluino  operators. 
The last column gives the number of components  associated to each one of them. 
The vector and tensor spin matrices  $V(\rho)$ and $T(\rho \sigma)$ are symmetric, 
thus forbidden when the field is Majorana.}

\vspace{1cm}

\input{tabla1.tex}

\end{center}
\end{table}

\newpage

\end {document}

%% file: section1.tex
\section{Introduction}

The non-perturbative aspects of the strongly interacting supersymmetric (SUSY) 
gauge theories were intensely investigated in the past~\cite{npSUSY} and
recently they have been the object of renewed 
interest~\cite{npSUSYrev,Seib-Witten}. 
These theories are interesting from a phenomenological point of view as
their non-perturbative properties might play a crucial role in 
the understanding of the SUSY breaking mechanism~\cite{npSUSY}.
However, besides the relevant phenomenological implications, the SUSY
gauge theories have intrinsic importance as their very nature
allows the calculation of some exact non-perturbative 
solutions~\cite{npSUSYrev,Seib-Witten}.

In this paper we concentrate our attention on the most simple SUSY gauge 
theory: the pure $\NS=1$ SUSY Yang-Mills (SYM) with $\SUN$ gauge group.
This contains the purely gluonic action, plus one flavour of Majorana 
fermions in the adjoint representation of the colour group.
It is believed that this theory is characterised by
the same non-perturbative phenomena as QCD:  colour 
confinement and chiral symmetry breaking~\cite{VY}.
Nevertheless, noticeable differences between the
SYM theory and  QCD appear to exist even at the fundamental level. 
Due to supersymmetry there is indeed a new anomalous SUSY current 
which belongs to the same supermultiplet together with
the anomalous chiral and energy-momentum tensor currents~\cite{npSUSYrev}.

The fundamental question of the breaking of the supersymmetry in $\NS=1$ SYM theory
was addressed in~\cite{VY,Windex}. According to the general
argument of the Witten index~\cite{Windex} or 
the Veneziano-Yankielowicz low energy effective theory~\cite{VY}, 
the supersymmetry does not break.
Nevertheless, here the chiral symmetry breaks and the
gluino condensate acquires a vacuum expectation value~\cite{VY,SUSYLOW}.
However, the authors of Ref.~\cite{Minkowski} argue that in this
theory the spontaneous breaking of chiral symmetry implies the spontaneous breaking of supersymmetry
due to non-perturbative effects.

From~\cite{VY,SUSYLOW} 
we learn that the low energy supermultiplet contains three 
degenerate massive colourless bound states: a scalar, a pseudoscalar 
and a fermion field (where the appearance of the fermion field in the 
low-energy supermultiplet is a consequence of the colour 
adjoint representation for the gluino).
Moreover, there is no pseudo-goldstone boson (or pion) associated 
with the chiral symmetry breaking, as the latter is broken by the anomaly.

Some time ago Curci and Veneziano~\cite{cv} suggested that the 
SYM theories can be studied non-perturbatively on the lattice 
by using numerical Monte Carlo simulations.
This is clearly analogous to the approach used in QCD theory.
They argued ~\cite{cv} that,  
even if the lattice breaks explicitly supersymmetry,
it is possible to recover the SUSY and chiral Ward identities on the 
continuum limit. 

Recently two different collaborations~\cite{montvay}-\cite{munster}
studied non-perturbatively
on the lattice the spectrum of the SYM ($\NS=1$) theory following the
guidelines suggested in~\cite{cv}.
In~\cite{andrea}, because of the
limitations deriving from the use of computing resources,
the quenched approximation was used to study the spectrum.
This approximation consists in neglecting the internal gluino loops.
In other words, in the correlation functions of fermion composite operators,
the fermion determinant is not included.
In  SYM theories, if general arguments are taken into
account~\cite{VY},  the quenched approximation cannot be justified on the
basis of large $\NC$ dominance, since  gluinos are in the adjoint 
representation of the colour group.
However, in~\cite{andrea} it is observed that,
within the statistical errors of the Monte Carlo simulation,
the spectrum seems to show no deviations from the  
supersymmetry expectations under the approximations
considered there.
In connection with this result, in~\cite{qsusy}, 
by means of an effective lagrangian approach,
the splitting in the low energy supermultiplet induced by 
the quenched approximation has been analysed,  and it is argued that
the  splittings  from the supersymmetry predictions are   small. 

In any case,  in~\cite{dg,munster}  unquenched results
for an SU(2) SYM gauge theory were obtained by using numerical
simulations with dynamical gluinos. It is likely that new results
in this field will quickly follow.

In this paper we study the spectrum of the SYM $(\NS=1)$ on the lattice
at strong coupling and in the large $\NC$ limit.
The lattice strong coupling expansion technique (see \cite{CreutzRothe} and references therein)
has been extensively 
used as an analytical probe to test qualitative properties both of the continuum and of the 
lattice theory by itself.
Up to now, however, the main part of the results refers to QCD, and --to our knowledge-- no
work deals with a supersymmetrised version of it. In detailed computations the strong coupling
expansion of $\SUN$ theories is frequently  combined with the large $\NC$
one. 

Our reference comes from works in which issues such as the phase structure of QCD or the computation
of meson and baryon masses are addressed 
\cite{BreKlub}-\cite{Froh}.
The most frequent computational frameworks split into two types:

\begin{itemize}

\item {\it Effective actions:} The Wilson-Dirac lattice action is considered
at large $\NC$ and small $\beta$.
The large $\NC$ expansion can be recognised as a saddle-point expansion of the 
gauge functional integral, 
previously simplified by the $\beta \to 0$ limit\cite{BreKlub}-\cite{KawSmit}.  
The method allows the study of the large and small hopping parameter regions
at the expense of introducing assumptions on the form of the saddle points.
The main  disadvantage of this method for our purposes is that the one matrix 
integration formulae, a basic ingredient of the construction, are not available for
matrices in the adjoint representation, and the generalization does  not seem straightforward.

\item {\it Path resummation:} The fermion matrix $\mathbf{M}$ is inverted by using
the standard hopping  parameter expansion \cite{CreutzRothe,Kawam,Froh}. This allows to compute
propagators $\mathbf{M^{-1}}$ in terms of sums over paths on
 the lattice (random walks); the objects to be summed are traces of products of spin
 matrices  $(r\mathbf{I} \pm \gamma_{\mu})$,  $r$ being the Wilson parameter.
We will see that provided the hopping parameter $\kappa$ is
small enough and some constraints on the parameters are imposed,
no assumptions are required to perform this computation in the
$\NC \to \infty, \, \beta \to 0$ limit.
In some of the previous references the analysis has been performed 
only for the case $r=1$, where the structure 
of the series is considerably simplified \cite{Kawam}. 
The main difficulty of path combinatorics arising for $r\ne 1$
was addressed in Ref.~\cite{resum}. In Ref.~\cite{combi} 
an independent derivation of the main resummation
formulas for $r\ne 1$ is given.
\end{itemize}

In our case, it will be shown that the hopping parameter expansion in terms
of random walks is valid, with slight changes, for the case of gauge fields in
the adjoint representation. Thus, using the formulas of Ref.~\cite{combi} we
will be able to give expressions for propagators and masses for any value of
$r$ and small enough $\kappa$.
We considered that keeping $r$ arbitrary could be
very important. This allows the possibility of searching for multicritical
points, where several lattice masses  vanish.
In particular, we investigated whether one could find a critical point
in the $\kappa-r$ plane where the massless modes would form a 
supermultiplet. This would signal a possible candidate for a supersymmetric 
continuum limit, in the spirit of the chiral restoration of ordinary QCD. 
For example, if a low energy theory of the type described in~\cite{cv}
would take place, one would find  a massless pseudoscalar meson, 
together with a scalar meson and a fermion. We should warn the reader that by 
massless we refer to the vanishing of the lattice masses, not necessarily the 
renormalised physical masses. However, in our approximations the 
 diagrams contributing to the anomaly, and giving mass to the pseudoscalar, 
 are subleading, indicating that  the physical masses are also zero, as
predicted for a  Goldstone boson associated to the spontaneous breaking of chiral
symmetry.  This situation contrasts with what one obtains in perturbation 
theory and constitutes one of the most salient features of our result.

In addition to the path resummation formulae, we also need to analyse
the behaviour in the large $\NC$ limit of integrals over the group of 
products of matrices in the adjoint representation. These are
studied in~\cite{largeN}.  In this paper we will only need to know the order
in $\NC$ of certain integrals. Given this, the actual leading  order 
results are very simple.

To increase the usefulness of our paper, many of our formulas will be
given  for  arbitrary  space-time dimensionality $d$. In addition, we will 
indicate the necessary changes to make the formulae valid  for matter fields 
in the fundamental representation and for Dirac, rather than Majorana, fermions. 
Specific attention will be paid, however, to the supersymmetric cases of $d=3,4$.

The paper is organised as follows: in section 2 we present the
formalism and general formulae for the expectation values and correlation functions
of two and three gluino operators in the strong coupling large $\NC$ limit.
In section 3 we analyse the specific channels and present the results for
the spectrum in the meson and fermion sector.
Finally in section 4 we summarise our conclusions, discuss the physical
implications and  explain the prospects for possible future extensions of 
our results.
The paper is completed by collecting in the Appendix the terminology and
formulae on lattice paths that we will use along the text.

%% file: section2.tex
\section{General formalism}
We begin by fixing our notation and terminology. Let us consider  a
$d$-dimensional hypercubic lattice ${\cal L} \equiv Z^d$. We introduce an 
index set $I$ having $2d$ elements.
To any direction in space-time we associate two indices $\mu$ and $\bar{\mu}$
corresponding to the two opposite senses (forward and backward) associated to
each direction $\mu$. Given an element $\alpha \in I $, the element
$\bar{\alpha}$ labels the one with  reverse orientation  ($\bar{\bar{\mu}}= \mu$).
To any element $\alpha \in I $, we can associate a vector $\GV{\alpha}$ as
follows:
\be
\GV{\mu}=e_{\mu}\ \ \ \ \ \ \ \ \GV{\bar{\mu}}=-e_{\mu}\ \ ,
 \ee
 with $e_{\mu}$ the unit vector in the $\mu$ direction (the lattice spacing is
set to 1). In the Appendix we 
 give some additional results and terminology that we will be using in the
 following sections.
 
Let us now  write  down the lattice version of  the SUSY 
Yang-Mills action:
\be
S = \beta S_g + \frac{1}{2} \Psi_i \Psi_j \bM_{i j }\ \ ,  
\ee
where $\beta S_g$ is the pure gauge part and $\Psi_i$ is a Grassman variable
representing  the field of a   Majorana fermion.
The index $i$ is a short form for the three
indices $n$, $a$ and $A$. The index $n$
specifies  a point in the $d$-dimensional hypercubic lattice ${\cal L} $.
The index $a$ takes $\NC^2-1$ values corresponding to the dimension of
the Lie Algebra of $\SUN$. The index $A$ is a spinorial index taking
$2^{[d/2]}$ values (the symbol $[x]$ stands for integer part of $x$).
Without much problem, but at the expense of breaking 
supersymmetry, we could add a flavour index
ranging over a finite number of values.  The matrix $\bM$ must be
antisymmetric and its form is given by 
\bea
& \bM = C\ \cM   \\
& \cM = \left( \mathbf{I} - \sum_{\alpha \in I} 
\mathbf{\Delta}_{\alpha}\right)\ \ , 
\eea
where $\mathbf{I}$ is the unit matrix and
$C_{A B}$ is the charge conjugation matrix, satisfying:
\bea
\label{cond1}
&\gamma_{\mu}^t\, C = - C\, \gamma_{\mu}  \\
\label{cond2}
& C^t= -C \, ,
\eea
where the superscript $t$ stands for transpose.

Finally the matrix $\mathbf{\Delta}_{\alpha}$ is given by:
\be
(\mathbf{\Delta}_{\alpha})_{i j }= \kappa\,  \delta_{m\, n+\GV{\alpha}}\
U_{\alpha}^{a b}(n)
\,(r \mathbf{I} - \gamma_{\alpha})_{A B} \ \ ,
\ee
with $i=(n,a,A)$,  $j=(m,b,B)$; $\kappa$ is the hopping parameter, and $r$ is the Wilson parameter.
For $\alpha=\mu$,  $U_{\mu}(n)$ is the gauge field link
variable (which is in the adjoint representation) and $\gamma_{\mu}$ the
Dirac matrix, while for $\alpha=\bar{\mu}$
we have $U_{\bar{\mu}}(n)=U^t_{\mu}(n-e_{\mu})$ and
$\gamma_{\bar{\mu}}=-\gamma_{\mu}$. Notice that we have:
\be
\mathbf{\Delta}_{\alpha} \mathbf{\Delta}_{\bar{\alpha}} =
\kappa^2 (r^2-1)\, \mathbf{I} \,\,\, .
\ee
It is easy to show that given the conditions Eqs.
(\ref{cond1})-(\ref{cond2}), the matrix $\bM$ is indeed antisymmetric.

Customarily, the  value of $r$ is taken in the interval $[0,1]$.
This follows from the requirement of  Osterwalder-Schrader positivity.
However, at strong coupling this is not a necessary condition for the  physical
correlation functions to admit an analytical continuation to unitary 
Minkowski Green functions. For that reason we will be working for  arbitrary $r$ 
and comment at the relevant places whether the $p$-gluino  correlation 
functions satisfy the positivity conditions.

The constraint imposed by the existence of the  matrix $C$, together
with the needed matching of fermionic and bosonic degrees of freedom,
makes the naive continuum limit of the above lagrangian 
supersymmetric ($\NS=1$) in $d=3,4$.
The same lagrangian is supersymmetric in $d=10$ if $\Psi$ is
a Majorana-Weyl field, and in $d=6$ if $\Psi$ is a Weyl spinor.
However, the requirement of Weyl character conflicts with the well-known
absence of chirality on the lattice, which prevents us from directly writing a lattice
version of the required lagrangian (indeed, our Wilson-type action breaks
chirality explicitly). On the other hand, the general argument by Curci and Veneziano
\cite{cv} linking the recovery of chirality and supersymmetry could play a role
in the interpretation of these cases, as of course does in the (more transparent) $d=3,4$ as well.

Having in mind this caveats regarding the interpretation of the results,
in the rest of our exposition we will try
to work as much as possible in arbitrary dimension $d$ without specifying more.
This has the advantage that it allows our formulae to be used with little
changes by other researchers with other physical interests. For example, links can
be established with the technique of $1/d$ expansions~\cite{oovd}.

We will concentrate upon gauge invariant operators  of the form:
\be
\bO_i(x)=\Psi_{A_1}^{a_1}(x) \ldots \Psi_{A_p}^{a_p}(x)\,
\cS_i^{A_1 \ldots A_p}\, \cC_i^{a_1 \ldots a_p} \,\,\, ,
\label{defopers}
\ee
where $\cC_i^{a_1 \ldots a_p}$ is an invariant colour tensor and
$\cS_i^{A_1 \ldots A_p}$ a spin tensor. The index $i$ labels different
possibilities for the definition. The restriction to operators obtained  by 
multiplying gluino fields at the same lattice point (ultralocal operators)
will not affect our spectrum results which will be general. In particular, 
this affects the composite fermion operator  which belongs to the same 
supermultiplet as the scalar and pseudoscalar:
\be 
\tilde{\chi}^A(x)=\frac{1}{2}(\sigma_{\mu \nu})^{AB}F_{\mu \nu}^a(x) \Psi^a_B(x) \ \ ,
\label{contferm}
\ee
where in this formula the fermion fields $\tilde{\chi} \, , \, \Psi$ and the Yang-Mills
field $F_{\mu \nu}$ live in the continuum spacetime, and
$\sigma_{\mu \nu}=\frac{1}{2} \lbrack \gamma_{\mu},\gamma_{\nu} \rbrack$.
Later on, we will argue that its corresponding minimal mass will be contained 
within  the set of masses corresponding to 3-gluino operators.

Note that the operators in Eq.~\ref{defopers} are non-vanishing only when the 
gluino fields are combined in a completely antisymmetric way. By putting this
together with the requirement of
$\cC_i$ being an invariant tensor, we arrive at certain constraints on the
possible operators; for example, in the case of meson-like operators ($p=2$)
it turns out that, necessarily, $\cC_i=\mathbf{I}$ and $\cS_i$ is an antisymmetric matrix.
As an example,
Table~1 displays a complete basis of 2-gluino operators in $d=4$.

In this paper we will be interested in computing the 
expectation values of these operators and products of these operators:
\bea
\label{expec1}
\langle\bO_i(x)\rangle &=& \frac{1}{Z}\, \prod_{n \in {\cal L}}\left( 
\int d\Psi(n)\, \prod_{\mu} \int dU_{\mu}(n)\right)
\bO_i(x)\, e^{-S}
\\
G_{i j }(x-y) &\equiv&  \langle\bO_i(x) \bO_j(y)\rangle = \nonumber \\
\label{expec2}
&&\frac{1}{Z}\, 
\prod_{n \in {\cal L}}\left(\int d\Psi(n)\, \prod_{\mu} \int
dU_{\mu}(n)\right)\, \bO_i(x)\, \bO_j(y)\, e^{-S}
\eea
at strong coupling.
We will be able to accomplish this goal  for $\beta=0$, and in the large
$\NC$ limit. Corrections to the formulae in 
powers of $\beta$ and $\frac{1}{\NC}$
are  in principle feasible and will be considered elsewhere. 

As usual, one
can explicitly integrate out the fermions in 
Eq.~(\ref{expec1})-(\ref{expec2}). The main formula that one uses is:
\be
\label{ferint}
 \prod_{i}\left( \int  d\Psi_i\right)\, \exp\{-\frac{1}{2} \Psi_i \Psi_j\,
\bM_{i j } + J_i \Psi_i \} = Pf(\bM) \exp\{-\frac{1}{2} J_i  J_j
(\cM^{-1} C^{-1})_{i j} \} \,\,\, ,
\ee
where $Pf(\bM)$ stands for the Pfaffian of the matrix $\bM$.
The square of the Pfaffian is the determinant, so that up to a sign
we can write:
\be
\label{PF}
Pf(\bM)= \sqrt{\det(C) \det(\cM)} = \exp\{\frac{1}{2}
Tr(\log(\cM)) \} \,\,\, ,
\ee
where we have used the standard exponential representation of a determinant,
and the fact that we can choose $\sqrt{\det(C)}=1$. Concerning the sign, it
is clear that what matters is the  relative sign  for different values of
the gauge field configuration. However, for very small $\kappa$, the matrix
$\cM$ is close to the unit matrix and our representation (Eq.~(\ref{PF}))  is valid.
Problems can only arise when one of the eigenvalues of $\cM$ crosses
zero. Using the Schwartz inequality and our expression for $\cM$, one
can conclude that  this problem never occurs provided $|\kappa| < \frac{1}{2
d\, (|r|+1)}$.  

\vspace{1cm}

What we will do now  is to expand the quantities entering in
Eq.~(\ref{ferint})  as a sum of paths.
Using the terminology of the Appendix, we obtain:
\bea
\label{form1}
&(\cM^{-1}(x,y))^{a b}_{A B}=\sum_{\gamma \in \cS(x \rightarrow
y)} W^{a b}(\gamma)\,
\Gamma_{A B}(\gamma)\\
\label{form2}
&Pf(\bM)=\exp\{\frac{1}{2}\sum_{x \in {\cal L}} \sum_{L=1}^{\infty}
\sum_{\gamma \in \cS_L(x \rightarrow x)} \frac{1}{L}\, Tr(W(\gamma))\, Tr(\Gamma(\gamma))\} \,\,\, ,
\eea
where $x,y$ are lattice points, $ W(\gamma)$ is the path ordered product (along the path $\gamma$) of the gauge
field link variables, and
$\Gamma(\gamma)$ denotes the appropriate product of the spin matrices:
\be
\Gamma(\gamma\equiv (x,\vec{\alpha}))= \kappa^L (r -\gamma_{\alpha_1})
\cdots  (r -\gamma_{\alpha_L}) \ .
\ee

Let us  first consider  2-body operators:
\bea
\label{twobody}
&\bO_i(x)= \cS_i^{A B }\, \Psi^a_A(x) \,\Psi^a_B(x) = 
\overline{\Psi}(x) \widehat{S}_i \Psi(x) \\
&\mbox{with   }\widehat{S}_i = C^{-1} \cS_i \, .
\label{defshat}
\eea
The second expression in (\ref{twobody}) has  the same form as for Dirac
fermions, the difference being that for Majoranas we have  the relation
$\overline{\Psi}=\Psi^t C$. Notice that $\cS_i$ is an antisymmetric matrix. 
Using the formula~(\ref{ferint}) and expanding by using Wick's theorem we obtain:    
\bea 
\langle\mathbf{O_i}(x)\rangle &=& - \frac{1}{\langle Pf(\bM)\rangle_g} \langle Tr \left( \cM^{-1}(x,x)\,
C^{-1}\, \cS_i \right)
\,Pf(\bM)\rangle_g
\label{opre}
\\
\nonumber
G_{i j }(x-y)&=& \frac{1}{\langle Pf(\bM)\rangle_g} \langle Pf(\bM) \times \\
\label{gpre}
&&\lbrack - 2\, Tr\left( 
  C^{-1}\, \cS_i\cM^{-1}(x,y)\, C^{-1}\, {\cS}_j  
\cM^{-1}(y,x) \right)  \\
&&+ Tr\left( 
 \cM^{-1}(x,x)\, C^{-1}\, \cS_i \right)
 Tr\left(  \cM^{-1}(y,y)\, C^{-1}\,
\cS_j \right) \rbrack \rangle_g \,\,\, , \nonumber
\eea
where $Tr$ denotes the trace over colour and spin indices and
the symbol $\langle\ \rangle_g$ denotes expectation value with respect
to the pure gauge action part. The first  term on the right hand side of
Eq.~(\ref{gpre}) represents the  so-called {\em OZI} contribution, while
the second term contains the disconnected and non-OZI contributions \cite{cv}.
Now using the formulae~(\ref{form1})-(\ref{form2}) we can write back 
expressions~(\ref{opre})-(\ref{gpre}) in terms of products of sums of paths.

Up to now the expressions have been completely general, but now we will consider
the simplification arising from considering $\beta=0$ and $\NC \rightarrow
\infty$. What  we need to know is  the expectation value of the product of
traces of Wilson loops in this limit.  The main result that we will
use~\cite{largeN} is that for a collection of closed  paths $\{ \gamma_1, \ldots,
\gamma_s \}$ which are not pure spikes (see the Appendix) we have:
\be
\label{largeNform}
\langle\, Tr(W(\gamma_1))\, \ldots   Tr(W(\gamma_s))\,\rangle_g = O(1)\ .
\ee
This is true both for connected and disconnected expectation values. However,
for a pure spike path $W(\gamma)=\mathbf{I}$, and therefore its trace is
$\NC^2-1$. We see that the connected correlation between a pure spike Wilson
loop and any other operator vanishes. Our first conclusion is then that, at
leading order in $\NC$, the factor $Pf(\bM)$ cancels between numerator
and denominator. This  is precisely the quenched approximation, which turns
out to be exact in this limit. In principle, the result is surprising since
fermions carry indices  taking
$\NC^2-1$ values   (as gluons),  and the usual arguments why fermion loops are
subleading in the fundamental representation do not apply here. This adds to
the results obtained by other methods~\cite{andrea,qsusy},  pointing towards
the fact that the deviations  introduced by the quenched
approximation are not too large.    

Other conclusion that follows from formula (\ref{largeNform}) is the
suppression in the large $\NC$ limit of the non-OZI contributions to
the connected correlation functions. This includes the mass induced by
the anomaly on the pseudoscalar Goldstone boson. We should, hence, expect
such a massless state signaling the recovery of chiral symmetry and its
spontaneous breaking. As mentioned in the Introduction this result,
as well as the exactness of the quenched approximation, contrasts with the
behaviour obtained in perturbation theory to leading order in $\NC$.

Our results have been obtained at $\beta=0$. However, one can
introduce the pure gauge action  in various  ways having the same naive
continuum limit.  If one chooses to write the Wilson action in the adjoint
representation of the group,  formula~(\ref{largeNform}) implies that to
the order we are working (and in the small $\beta$ region) there are no
corrections to any order in $\beta$. This is not the case if the customary 
fundamental representation Wilson action is chosen.

The conclusion of the previous paragraphs is that  to leading order in $\NC$ all 
that we have to take into account are closed loops which are  pure  spikes. For that 
purpose the results of Ref.\cite{combi} and collected in the Appendix are 
needed. 
Notice  that for a pure spike path
of length $L$, the spinor matrix is given by:
\be
\Gamma(\gamma)= (\kappa^2(r^2-1))^{L/2}\, \mathbf{I} 
\ .
\ee
Thus, using the formulae given in the Appendix we can conclude:
\bea
\nonumber
\langle\mathbf{O_i}(x)\rangle &=& -(\NC^2-1)\, F(0,\kappa^2(r^2-1))\,
Tr(  C^{-1}\, \cS_i)=\\
&& - (\NC^2-1)\, Tr(\widehat{S}_i)\, \frac{1}{1-\frac{2d}{2d-1} \xi} \ \ ,
\label{condeq}
\eea
with
\be
\label{xidef}
\xi \equiv \frac{1-\sqrt{1-4(2d-1)\kappa^2(r^2-1)}}{2} \ . 
\ee
This is the contribution of closed paths which are pure spikes. The
corrections coming from other paths are order $1$, and thus subleading.
The previous formulae are obtained by resummation of a series. The radius of 
convergence is given by the closest singularity. Thus the formulae are 
strictly speaking only valid in the  region $|\kappa^2(r^2-1)| < \frac{1}{4 (2d-1)}$.
It is possible that, by analytical continuation,  the
formulae could be valid in some points  beyond this region, such as for
larger negative $\kappa^2(r^2-1)$.  

Expression~(\ref{condeq}) is also valid for  Dirac
fermions. It is also valid if the fields (Dirac or Majorana) transform in the fundamental
representation of the colour group, provided $(\NC^2-1)$ is replaced by
$\NC$, the dimension of the representation in question.

Now we look at the correlation function of   two-fermion  operators. In this
case we have two factors of $\cM$ and  hence we have an expansion in
terms  of paths $\gamma$ going from $x$ to $y$, and paths $\gamma'$ going
from $y$ to $x$. Nevertheless, the integration over the gauge group forces
the combined path to be a pure spike path. To take this into account we
proceed as follows. We replace the summation over paths by a summation over
paths with no spikes, resumming all paths which have such a path as reduced
path. Thus, each term in the new expansion  corresponds to a reduced path
$\hat{\gamma}$  going from $x$ to $y$, and another one $\hat{\gamma}'$ that
returns to $x$. However, now the condition imposed by the integration over
the gauge group is simply  that  $\hat{\gamma}'$  is the reverse path of
$\hat{\gamma}$ (which we label $\hat{\gamma}^{-1}$).
In this way the double summation reduces to a single summation.  
Summing up all that we have just expressed in words, we can give the 
following formula for the connected correlation function:
\bea
\nonumber
 G_{i j}^{(conn.)}(x-y) &=&  -\eta~D_R \,
\sum_{L=0}^{\infty} \sum_{\hat{\gamma} \in \bar\cS_L(x \rightarrow y)}\,
(F(L,\kappa^2(r^2-1)))^2 \times \\
&& Tr\left( \Sh_i \Gamma(\hat{\gamma})\, \Sh_j \,
\Gamma(\hat{\gamma}^{-1})   \right) \, .
\eea
The previous formula has been written in a way which makes it valid for
Majorana ($\eta =2$)   or Dirac ($\eta = 1$) fermions. The symbol  $D_R$ stands 
for the dimension of the gauge group representation ($\NC^2-1$ for the adjoint
and $\NC$ for the fundamental).
Now we can use the  expression for $F(L,\kappa^2(r^2-1))$ and the formulae for
resumming over paths that are given in the Appendix, to conclude:
\be
G_{i j}^{(conn.)}(x-y) = R_2(\xi)\,
\prod_{\mu} (\int \frac{d\varphi_{\mu}}{2 \pi})\, e^{\imath \varphi (x-y)} 
  \langle \cS_i |\,\lbrack \Theta_2(\xi)\mathbf{I} - \tAtwo(\varphi)
\rbrack^{-1} \, \CTT^{-1}| \cS_j \rangle \ ,
\label{gijmeson}
\ee
where  we have:  
\bea
 R_2(x) &\equiv& -\eta~D_R \, \frac{1-\frac{2d-2}{2d-1}x}{1-\frac{2d}{2d-1}x}\\
\label{deftheta2}
\Theta_2(x) &\equiv& (1 -x)^2 +\frac{x^2}{(2d-1)} \\
\label{deftAtwo}
 \tAtwo(\varphi) &\equiv& \kappa^2\, \sum_{\alpha \in I}  e^{\imath
\varphi_{\alpha} }
 \rmga \otimes \rmga \\
 \CTT \, &\equiv& \, C \otimes C \, .
 \eea
$| \cS_i \rangle$ is just given by the matrix $\cS_i$, but considered  as  a
 $2^{[\frac{d}{2}]}\cdot 2^{[\frac{d}{2}]}$ dimensional vector. It is useful
to express  the matrix elements of   the
$2^{2 \lbrack \frac{d}{2}\rbrack} \times 2^{2\lbrack\frac{d}{2}\rbrack}$ matrix
$\tAtwo$ between $\cS_i$ states in terms of the matrices $\Sh_i$ defined in
Eq.~(\ref{defshat}). We have:
\be
\langle \cS_i |\, \tAtwo(\varphi) \, \CTT^{-1} \,| \cS_j \rangle = \kappa^2\, \sum_{\alpha \in I}  e^{\imath
\varphi_{\alpha} }
\, Tr\lbrack \Sh_i \rmga \Sh_j \rpga\rbrack \ .
\ee
With this interpretation,  formula~(\ref{gijmeson}) is valid for Dirac
fermions as well.  
 We will leave to the next section the evaluation of  this expression and 
 the study
of the properties of the resulting propagator.

\vspace{1cm}

We now turn our attention to 3-gluino fermion operators of the form:
\be
\label{threegops}
\bO_i(x)=\Psi^{a_1}_{A_1}(x) \Psi^{a_2}_{A_2}(x) \Psi^{a_3}_{A_3}(x) \cC_i^{a_1 a_2 a_3} \cS_i^{A_1 A_2 A_3} \ .
\ee
In this case there are
two possible invariant colour tensors: $d_{a b c}$ and $f_{a b c}$. The main result
that we will  need on the group integration at large $\NC$ is  the following:
given three  paths without spikes $\gamma_1$,  $\gamma_2$ and  $\gamma_3$, we have: 
\bea
\nonumber
\cC^{a b c}\,  \cC^{a' b' c'} \langle W^{a a'}(\gamma_1)W^{b b'}(\gamma_2) W^{c
c'}(\gamma_3) \rangle_{g} &=& \NC^3  \delta(\gamma_1=\gamma_2=\gamma_3) \\
&&+ \mbox{subleading terms} \ ,
\label{prev}
\eea
where $\cC^{a b c}$ is either $f$ or $d$ ( the antisymmetric and symmetric
$\SUN$ invariant tensors). The mixed terms $f-d$ are subleading
in $\NC$. Also subleading are   contributions in which the three paths are
non-equal. Using this expression and the formulae  derived in the Appendix,
it is possible to compute  the expectation values of products of
3-gluino operators. Now the Wick expansion gives a total of 6 terms (once the
$\NC$-subleading ones are discarded). These terms ensure that if the colour
invariant tensor is $f$ or $d$, the spin matrix $\cS_j$ can be chosen completely symmetric
or antisymmetric respectively, as required  by Fermi statistics. With this
choice  the 6 terms give the same contribution,
and the correlation function can be written as:
{\setlength\arraycolsep{1pt}
\bea
\label{tgluino}
\nonumber
\langle \bO_i(x) \bO_j(y) \rangle &=& 
R_3(\xi)\,
\prod_{\mu} (\int \frac{d\varphi_{\mu}}{2 \pi})\, e^{\imath \varphi (x-y)}\,
 \langle \cS_i |\,  \lbrack \Theta_3(\xi)\mathbf{I} -
\tilde{\mathbf{A}}_3(\varphi) \rbrack^{-1} \CTE^{-1}\,| \cS_j \rangle \,\,\, , \\
\eea}
where now:
\bea
 R_3(x) &\equiv& - 6 \NC^3 \, \frac{(1-x)^3 -(\frac{x}{2d-1})^3
}{(1-\left( \frac{2d}{2d-1}\right) x)^3} 
\\
 \Theta_3(x) &\equiv& (1-x)^3 + \frac{x^3}{(2d-1)^2}
\\
  \tilde{\mathbf{A}}_3(\varphi) &\equiv& \kappa^3\, \sum_{\alpha \in I}  e^{\imath
\varphi_{\alpha} }
 \rmga \otimes \rmga \otimes (r -
\gamma_{\alpha})\\
\CTE &\equiv&  \, C \otimes C \otimes C \, . 
\eea
As before, the vector  $| \cS_i \rangle$ is the one constructed from the
corresponding spin matrix. Explicitly:
\bea
&\langle \cS_i |\, \tilde{\mathbf{A}}_3 (\varphi) \, \CTE^{-1} \,| \cS_j \rangle =
\kappa^3\, \sum_{\alpha \in I}  e^{\imath \varphi_{\alpha} } \times \\
&\, \cS_i^{A_1 A_2 A_3} (\rmga C^{-1})^{A_1 B_1} (\rmga C^{-1})^{A_2 B_2}
(\rmga C^{-1})^{A_3 B_3} \cS_j^{B_1 B_2 B_3} \,\,\, . \nonumber
\eea

The class of operators considered (Eq.~(\ref{threegops})) does not include the
lattice counterpart of that in Eq.~(\ref{contferm}). A possible candidate would
be:
\be
\chi^A(x)=\frac{1}{2}(\sigma_{\mu \nu})^{AB}{\cal P}_{\mu \nu}^{ab}(x)f^{abc} \Psi^c_B(x) \ \ ,
\label{ferlatop}
\ee
where ${\cal P}_{\mu \nu}(x)$ is an appropriate combination of adjoint plaquettes in the
$(\mu,\nu)$ plane whose naive
continuum limit is, up to a convenient multiplicative factor, the adjoint gauge field
$\mathbf{F}_{\mu \nu}(x)$.
This has the advantage of including
only gauge variables in the adjoint representation, allowing the use of our
integration formulas. However, by examining the strong coupling
large $\NC$ expansion of the correlation of 2 such operators, one easily 
realises that it is given by combinations of triple paths joining the two 
operators. Thus, we expect that the mass spectrum following from  3-gluino 
ultralocal operators would include also the states coupled to (\ref{ferlatop}).

To conclude we simply want to mention that in a similar way one can obtain
expressions for expectation values and correlations of $p$-gluino operators. No
additional difficulty arises, and the final expression looks just like
Eq.~(\ref{tgluino}) but with  corresponding functions and matrices $R_p$,
 $\Theta_p$, $\tilde{\mathbf{C}}_p$, and  $\tilde{\mathbf{A}}_p$. In particular:
\bea
\label{thetap}
 \Theta_p(x) &\equiv& (1-x)^p + \frac{x^p}{(2d-1)^{p-1}} \\
  \tilde{\mathbf{A}}_p(\varphi) &\equiv& \kappa^p\, \sum_{\alpha \in I}  e^{\imath
\varphi_{\alpha} }
\underbrace {\rmga \otimes \ldots \otimes \rmga}_p \, .
\label{atp}
\eea

%% file: section3.tex
\section{Explicit results on the propagators and spectra} 
In this section we will analyse the results on the expectation values 
presented in the previous one. Our main goal will be the extraction of
the spectrum of the theory  at $\beta=0$ and leading order in the $1/\NC$ expansion.

We  will first of all look at the expectation values of single 2-gluino operators.
Our main result is formula~(\ref{condeq}).  The only independent operator $\cS_i$ giving a
non-vanishing spinorial trace can be chosen as $\cS_i=C \, \mathbf{I}$. On
physical terms, it corresponds to a non-vanishing gluino scalar condensate for the
full range of values where our resummation is valid ($\frac{1-\sqrt{2}}{2} \le \xi \le
\frac{1}{2}$). Once the different normalisations of the fields and operators,
and the appropriate
colour factors are taken into
account, this expression  coincides with  the result given in
Ref.~\cite{KawSmit} for this expectation value when the fermion field is in the 
fundamental representation.

Our next step will be to analyse the results for the correlations of two
fermion operators $G_{i j}(x)$, and the corresponding ({\em meson}) spectrum. The
expression given in the preceding section for the correlation
(formula~(\ref{gijmeson})) requires the inversion of a $2^d \times 2^d$ matrix.
(For odd space-time dimensions one must replace $d$ by $2[\frac{d}{2}]$).
This matrix is  $\Theta_2(\xi)\mathbf{I}-\tAtwo(\varphi)$
defined in Eqs.~(\ref{deftheta2},\ref{deftAtwo}). Thus, in even space-time dimensions, it is convenient for
the study to choose as a
basis of the  $2^{d}$-dimensional space  of meson operators, those
corresponding  to
$\widehat{S}_i$ being the standard basis of the 
$d$-dimensional Clifford Algebra, for which we adopt the writing:
\be 
\label{defgamma}
\Sh(n_{\mu})
 \equiv e^{\imath \delta(n_{\mu})}\, \gamma_0^{n_0} \cdots
\gamma_{d-1}^{n_{d-1}} \,\,\, ,
\ee
where $\delta(n_{\mu})$ is an appropriate phase which we will choose equal
to zero in what follows.  Thus, one state of the basis is the scalar operator
$\Sh_S$ corresponding to the unit matrix, other elements are the
vector operators $\Sh_{V(\mu)}$ corresponding to the gamma matrices
$\gamma_{\mu}$, and so on.   

For odd space-time dimensions, one could also consider the operators
$\Sh(n_{\mu})$ associated to the standard basis of the Clifford algebra (which is basically
the same as for dimension $d-1$), but one must take into 
account that they are not independent. They are constrained by the identity:
\be
\label{constraint1}
\gamma_0\cdots \gamma_{d-1}=K_d \, \mathbf{I} \,\,\, ,
\ee
where $K_d$ is a phase depending on the space-time dimension. 

Going back to the even-dimensional case,  we have to express the matrix elements of
the matrix $\tAtwo(\varphi)$ within this basis. To do so it is
convenient to view the $2^{d}$-dimensional space in question as the Fock
space of a system of fermions: the {\em gamma-fermions}. Each integer
 $n_{\mu}\in \{0,1\}$ entering Eq.~(\ref{defgamma}) can be interpreted as the occupation number of
 the state $\mu$. It is convenient to add an additional one-particle
 state labelled `$-1$' whose usefulness will be clear in what follows.
 Thus, in the standard second quantization notation we can write:
 \be
 \Sh(n_{\mu}) \equiv |n_{-1},n_0,\ldots n_{d-1}\rangle \,\,\, ,
 \ee
with $\Sh(n_{\mu})$ defined in (\ref{defgamma}).
The extra occupation number $n_{-1}$ is fixed to be a {\em parity bit}
state, taking the value $1$ when the total number of gamma-fermions in the other
states
is odd and the value $0$ if it is even. Hence, in both cases, this imposes
the constraint that  the total number of fermions must be even. 
With this convention it is possible to express the
matrix $\tAtwo(\varphi)$ in terms of creation and annihilation
operators of these fermions as follows:
\bea
\label{A2q}
\nonumber
&\tAtwo(\varphi)= \kappa^2( 2\tilde{\sigma}(r^2-1 +2 a^+_{-1}a_{-1}) -2(4
a^+_{-1}a_{-1} -2) \sum_{\mu=0}^{d-1} \cos(\varphi_{\mu})a^+_{\mu}a_{\mu} - \\
\label{gammafer}
&-4  \imath r
\sum_{\mu=0}^{d-1} \sin(\varphi_{\mu})(a^+_{\mu}a_{-1} + a^+_{-1}a_{\mu}
) )\,\,\, , \\
&
\nonumber
\mbox{ with }\ \  \tilde{\sigma}=\sum_{\mu=0}^{d-1} \cos(\varphi_{\mu}) \,\,\, .
 \eea
We see that the operator conserves the number of gamma-fermions. Hence, each
 even   number of gamma-fermions $2p$ characterises a block in which
 $\tAtwo(\varphi)$ can be diagonalised or inverted. Within each block one has
 two subspaces corresponding to $n_{-1}=1$ and  $n_{-1}=0$, which correspond
to  the product of $2p-1$ and $2p$ gamma matrices respectively.
The matrix  $\tAtwo(\varphi)$ mixes these 2
 subspaces.  

Let us clarify the previous formulae  by looking at a few examples. If
$p=0$,  we have the gamma-fermionic Fock vacuum state, which is an eigenstate of
$\tAtwo(\varphi)$ with eigenvalue $2\tilde{\sigma}\kappa^2 (r^2-1)$. This
state is precisely the scalar meson operator.
Next, we  consider the space of $2p=2$ gamma-fermions. The subsector
$n_{-1}=1$ corresponds to the vector operators ($\Sh_{V(\mu)}=\gamma_{\mu}$)
and the $n_{-1}=0$ to tensor states ($\Sh_{T(\mu \nu)}=\gamma_{\mu} \gamma_{\nu}$).
The conclusion is that vector and tensor states mix between themselves but
not with other states. Considering  the space  of $2p$ gamma-fermions, we
conclude that the operators associated to the product of $2p-1$ Dirac gamma
matrices mix with those involving $2p$ gamma matrices, but with no other
states.  The inversion or diagonalisation problem has been considerably 
simplified with this technique. This is the generalisation of the  block 
structure found by previous authors studying  QCD at strong coupling \cite{Kawam}. 

In the previous analysis we have not taken into account
the restriction imposed by the fact that our gluinos are Majorana. As
mentioned before, in this case if the operators involve the product of the
gluino fields at the same point (ultralocal operators), the matrices
$\cS_i$ can be chosen
antisymmetric. This restriction translates in  our language into
$p$ being an even number: the number of gamma fermions  must be a multiple
of 4. Thus, for instance, the only relevant blocks in $d=4$ for this ultralocal
operators are the scalar singlet and the $p=2$ containing the pseudoscalar
($\gamma_5$) and the axial vector ($\gamma_5 \gamma_{\mu}$).  

We now introduce an  important symmetry of the operator given in
(\ref{A2q}). This is the unitary transformation $\mathbf{C}$ related to the
{\em charge conjugation} of gamma-fermions:
\bea
& \mathbf{C} a^+_{\mu} \mathbf{C}^{\dagger} = a_{\mu} \\
&\mathbf{C} a^+_{-1} \mathbf{C}^{\dagger} = -a_{-1} \\
&\mathbf{C} |0\rangle= K'_d \, |1,\ldots,1\rangle \,\,\, .
\eea
Up to a phase the operation exchanges occupied by empty for all states.
One can easily see that the previous transformation commutes with the
operator $\tAtwo(\varphi)$ given in   expression (\ref{A2q}). Notice, that
with this change a state with $2p$ gamma-fermions changes into one with $(d+1-2p)$,
which for even space-time dimensions ($d=2s$) is an odd number. We see that the spaces
with an odd number of gamma fermions are useful after all. Thus, there is a
hierarchy  of complexity in the Fock space of gamma-fermions as the number
grows. The Fock vacuum corresponding to the scalar operator is an eigenstate
of $\tAtwo(\varphi)$. Next, comes the 1-particle space, which
through charge conjugation  of gamma-fermions corresponds to the operator
involving the matrix $\overline{\gamma}\equiv \gamma_0\cdots \gamma_{d-1}$ and
$\overline{\gamma}\gamma_{\mu}$. In what follows we will proceed to
invert the matrix $\Theta_2(\xi)\mathbf{I}-\tAtwo(\varphi)$ and obtain the
propagator for  these simplest cases.

Before proceeding to the inversion, let us comment about the necessary
changes to deal with an odd-space time dimension $d$. In this case we might
introduce gamma-fermions as well, but due to the constraint
(\ref{constraint1}) there are actually 2 states corresponding to the same
operator. However, with a bit of  effort one can show that the two states are
precisely the 2 states that are mapped by the transformation $\mathbf{C}$,
provided $K'_d$ is chosen equal to $K_d^*$ entering in Eq.~(\ref{constraint1}).
With this choice, one can see that for odd space-time dimension
expression (\ref{A2q}) remains valid, but that the {\em physical space} of meson
operators has to be identified with the subspace of the gamma-fermion Fock
space which is invariant under $\mathbf{C}$ and has an even number of
gamma-fermions.  With this in mind all that follows can be applied to even 
and odd space-time dimensions equally.

Now we consider  the scalar state (corresponding to the unit matrix) first.
Indeed, to comply with the normalisation chosen for the $d=4$ case in Table~1,
we take  $\Sh_S = 2^{-\frac{1}{2}\lbrack\frac{d}{2}\rbrack}\mathbf{I}$.
The matrix $\tAtwo(\varphi)$ reduces here to the constant $2 \tilde{\sigma}
\kappa^2 (r^2-1)$. Then we can directly write the momentum-space propagator explicitly:
\be
\widehat{G}_{SS}(\varphi) = \frac{H(\xi)}{\Phi_2(\xi)-\sum_{\mu=0}^{d-1} \cos({\varphi_{\mu}})}\no \,\,\, ,
\ee
where we have defined the following functions:
\bea
\label{defhphi2}
H(x)&=&\frac{-\eta~D_R(2d-1)\left( 1+ \frac{2d-2}{2d-1}x \right)}
{2x(1-x)\left( 1 - \frac{2d}{2d-1}x \right)}\no
\Phi_2(x)&=&\frac{(1-x)^2(2d-1)+x^2}{2x(1-x)} \,\,\, ,
\eea
and $\eta,~\xi$ and $D_R$ are the ones defined in section 2.
The function $\Phi_2$ is decreasing for all $\xi \ne 0$ in
the convergence interval,
and in addition satisfies the following  properties:
{\setlength \arraycolsep{2pt}
\bea
\Phi_2(1/2) &=& d \nonumber \\
\xi \in (0,1/2) &\Rightarrow& \Phi_2(\xi) > d \nonumber \\
\xi <0 &\Rightarrow& \Phi_2(\xi) < -d \,\,\, . \nonumber
\eea
}

Next, for even space-time dimension,  we proceed to study the space of 1 gamma-fermion. As mentioned
previously it corresponds to the matrix $\overline{\gamma}$ ($\gamma_5$ in 4
dimensions) and  $\overline{\gamma}\gamma_{\mu}$. Expression (\ref{A2q})
reduces in this case to:
\be
\tAtwo(\varphi)= 2 \kappa^2 (r^2-1) \tilde{\sigma} + \kappa^2 \sum_{\alpha,\beta=-1}^{d-1} T_{\alpha \beta}\,
a^+_{\alpha} \, a_{\beta} \,\,\, ,
\ee
where 
\bea
\nonumber & T_{-1 \,-1}= 4 \tilde{\sigma}\\
 & T_{-1\,\mu}=T_{\mu \,-1}= 4 \imath r \sin(\varphi_{\mu})\\
\nonumber & T_{\mu \,\nu}= 4 \, \cos(\varphi_{\mu}) \delta_{\mu \nu} \,\,\, . 
\eea
Now the propagator can be obtained by inverting the matrix $\Theta_2
\mathbf{I} - T$. This can be done by making a non-unitary change of variables
which brings $T$ to a diagonal matrix up to a $2 \times 2$ block. In this
way, one obtains the expression for the propagator in momentum space $\widehat{G}^{(PA)}$
in the 1-gamma fermion sector, i.e.,  the axial vector and the pseudoscalar
block in $d=4$. 
To make a contact with the 
usual conventions, we change the normalisation of the  operators to  the forms 
(remember $d$ is even) $P=2^{-d/4}\bar{\gamma}$ and 
$A(\rho)=\imath 2^{-d/4}\gamma_{\rho}\bar{\gamma}$,
generalising again the 4-dimensional ones given in Table~1. 
The result is:
\bea
\widehat{G}_{PP}(\varphi)&=&H(\xi)\left( \alpha_d + \sum_{\mu=0}^{d-1} \frac{\beta_{\mu}^2}{\alpha_{\mu}} \right)^{-1} \nonumber \\
\widehat{G}_{PA(\rho)}(\varphi)&=&\widehat{G}_{A(\rho)P}(\varphi)=
-\frac{\beta_{\rho}}{\alpha_{\rho}}\widehat{G}_{PP}(\varphi) \nonumber \\
\widehat{G}_{A(\rho)A(\rho)}(\varphi)&=& \frac{1}{\alpha_{\rho}}\left( \alpha_d + \sum_{\mu=0,\mu \ne \rho}^{d-1}
 \frac{\beta_{\mu}^2}{\alpha_{\mu}} \right) \widehat{G}_{PP}(\varphi) \nonumber \\
\widehat{G}_{A(\rho)A(\sigma)}(\varphi)&=&\widehat{G}_{A(\sigma)A(\rho)}(\varphi)=
-\frac{\beta_{\rho}\beta_{\sigma}}{\alpha_{\rho}\alpha_{\sigma}} \widehat{G}_{PP}(\varphi)~~~,~\rho \ne \sigma \,\,\, ,
\label{propPA}
\eea
where $H$ and $\Phi_2$ are the quantities defined in Eqs.~(\ref{defhphi2}), and the functions
$\alpha_d,~\alpha_{\mu}$ and $\beta_{\mu}$ have the following expression:
\bea
\alpha_d&=&\Phi_2(\xi)-\frac{r^2+1}{r^2-1}\sum_{\mu=0}^{d-1}\cos({\varphi_{\mu}})\no
\alpha_{\mu}&=&\Phi_2(\xi)-\sum_{\rho=0}^{d-1}
\cos({\varphi_{\rho}})-\frac{2}{r^2-1}\cos({\varphi_{\mu}})\no
\beta_{\mu}&=&2\frac{r}{r^2-1}\sin({\varphi_{\mu}}) \,\,\, .
\eea
Obtaining momentum-space propagators for other blocks is feasible, but the
expressions become more and more complicated. Furthermore, for ultralocal
operators in $4$ dimensions, the previous propagators are the only
non-vanishing ones. Thus, we will focus in what follows on the analysis of
the meson spectrum.

\vskip 1cm

The lattice masses are the  minima of the lattice energies
as we vary the spatial momentum $\vec{\varphi}$. These minima can only occur at
{\em special momenta} $\vec{\varphi}=\vec{\varphi}^{(special)}$
($\varphi_{i}^{(special)}=0,\pi$). The advantage is that now 
$\sin(\varphi_i^{(special)})=0$, which simplifies expression~(\ref{A2q}) 
considerably.  
The procedure to obtain the masses is the following:
Extract  the eigenvalues of the matrix
$\Theta_2(\xi)\mathbf{I}-\tAtwo(\vec{\varphi}^{(special)})$, which are
functions of the temporal momentum $\varphi_0$. Then  determine $\varphi_0^{pole}$,
the (complex) value of $\varphi_0$ for
which the eigenvalue vanishes.  The lattice masses are now given by
$M=-log(|e^{\imath \varphi_0^{pole}}|)$. This coincides with the definition
of mass as the exponent controlling the decay of correlation functions at
long times.

Now let us proceed to obtain the eigenvalues. By looking
at expression~(\ref{A2q}), one sees that the matrix is diagonal except
for the term proportional to $(a^+_{0}a_{-1} + a^+_{-1}a_{0})$. Thus, the
occupation numbers of the spatial states $n_i$ are not changed by the operator.
Hence,  for fixed values of these numbers, the operator reduces to a 4 by 4
matrix: an operator acting on the two-state fermion system labelled by
$n_{-1}$ and $n_0$. Furthermore,  the states having $n_{-1}=n_0=0$ and
 $n_{-1}=n_0=1$ are eigenstates of the matrix $\tAtwo(\vec{\varphi}^{(special)})$. 
 The other two states are
 mixed, but finding the eigenvalues and eigenvectors is trivial, since it is
 a $2\times2$ matrix. We can summarise the results obtained in the following
formulae:
\bea
\nonumber
&\tAtwo(\vec{\varphi}^{(special)}) |0,0,\vec{n}\rangle= (2 \kappa^2 (r^2-1)\, \cos(\varphi_0) +2
\kappa^2 r^2 \sigma - 2 \kappa^2 \tau)\,   |0,0,\vec{n}\rangle \\
\nonumber
&\tAtwo(\vec{\varphi}^{(special)}) |1,1,\vec{n}\rangle= (2 \kappa^2 (r^2-1)\, \cos(\varphi_0) +2
\kappa^2 r^2 \sigma + 2 \kappa^2 \tau)\,   |1,1,\vec{n}\rangle \\
&\tAtwo(\vec{\varphi}^{(special)}) |\mbox{mixed } \pm,\vec{n}\rangle= \\
&2 \kappa^2( r^2 \sigma +(r^2 +1)\, \cos(\varphi_0) \pm \sqrt{\tau^2 - 4 r^2
\sin^2(\varphi_0)})
|\mbox{mixed } \pm,\vec{n}\rangle \,\,\, ,
\nonumber
\eea
where we have introduced the following notation:
\bea
&  \sigma=\sum_{i=1}^{d-1} \vartheta_i  \nonumber \\
&\tau = \sum_{i=1}^{d-1} (-1)^{n_i} \vartheta_i \nonumber \\
&\epsilon = \frac{1}{r^2-1} \label{defvaria} \\
&\theta= \frac{r^2+1}{r^2-1} \nonumber \\
&\vartheta_i = \cos(\varphi_i) \in \{-1,1\} \nonumber \,\,\, .
\eea

From these eigenvalues one can apply the previously described
procedure and obtain the  formulae for the masses
$M(n_{-1},n_0,n_1, \ldots  ,n_{d-1})$:
\bea
\label{mscal}
& \cosh(M(0, 0,\vec{n}))= |\Xi + \tau \epsilon| \\
\label{m11}
& \cosh(M(1, 1,\vec{n}))= |\Xi - \tau \epsilon| \\
&\label{massmix} \cosh(M(\mbox{mixed } \pm ,\vec{n}))=
\theta\, \Xi \mp \sqrt{(\theta^2 -1)(\Xi^2-1) +
\tau^2 \epsilon^2} \\
\nonumber & \mbox{ with } \ \ \Xi= \Phi_2(\xi) -r^2 \epsilon \sigma \,\,\, ,
\eea
where $\Phi_2(\xi)$ is defined in Eq.~(\ref{defhphi2}) and the remaining symbols
in (\ref{defvaria}).

Now let us discuss these results. All the dependence on the occupation
numbers $n_i$ lies in the quantity $\tau$. On the other hand, both $\Xi$ and
$\tau$ depend on the choice $\varphi_i=0,\pi$. From the definition of
$\tau$ one sees that  its  maximum positive value  is obtained whenever all
states having $\vartheta_i=1$ are empty and those with  $\vartheta_i=-1$
occupied. The maximum negative value is attained in the opposite situation.
In both cases, the maximal absolute value is the same: $d-1$. Let us now
comment briefly on the main features of the spectrum formulae:

\subsection*{$n_{-1}=n_0$ sectors}
This sector contains the scalar state corresponding to $n_{-1}=n_0=n_i=0$, 
whose mass is given by:
\be
\cosh(M_S)= |\Phi_2(\xi)-\sigma| \,\,\, .
\ee
For $|r| > 1$ ($\xi >0$) the scalar ground state corresponds to zero momentum
$\vec{\varphi}=\vec{0}$. For $|r| <  1$ ($\xi < 0$) it corresponds to the
{\em doubler} state $\varphi_i=\pi$. The state with minimum mass within these sectors
corresponds to $|\tau|=d-1$ and $\varphi_i=0$, which is the state associated to
the matrix $\gamma_0$. Its corresponding mass is:
\be
\cosh(M_V) = | \Phi_2(\xi)-\theta(d-1)| \,\,\, .
\ee

\subsection*{$n_{-1}+n_0=1$ sector}

Let us consider first even space-time dimensions. In that case this sector has
lighter states than the previous one. Notice that the
argument of the square root in Eq.~(\ref{massmix}) can be written as
$(\theta \, \Xi -1)^2 - ((\Xi-\theta)^2 -\tau^2 \epsilon^2)$. Thus, the lightest state
corresponds to the maximum value of $|\tau|=d-1$.  Furthermore, one can prove
that the mass decreases  with $|\Xi|$, and hence the lightest state corresponds
to $\varphi_i=0$. Combining this with the maximal  $\tau$ one sees that this
lightest state corresponds 
to $\bar{\gamma}$ and $\gamma_0\bar{\gamma}$
 ( $\gamma_5$ and $\gamma_0 \gamma_5$ in 4 dimensions). Its mass is
given by formula (\ref{massmix}) with $\Xi=\Phi_2(\xi)-(d-1)r^2\epsilon$ and
$|\tau|=d-1$. The corresponding critical line, 
where this lightest meson becomes massless,  marks the edge of  the
physical region. It also marks the boundary  of validity of our formulae. 
The equation for this critical line is given by:
\be
\label{critline}
\Phi_2(\xi)=d\, \theta \,\,\, . 
\ee
Along this line, the other  meson 
states have a positive definite mass. The only exception occurs for $\theta=1$ ($\epsilon=0$).
The latter is an interesting region, where all the meson masses are degenerate and
dependent only on $\xi$. All states become massless at $\xi=\frac{1}{2}$. Notice
that the region corresponds to the limit $r\rightarrow \infty$, $\kappa
\rightarrow 0$ with $\kappa r $ fixed. If we look at the expression of the
action in this limit we see that the normal Dirac term becomes negligible 
with respect to the Wilson term. We have then  essentially the masses 
corresponding to the scalar-gauge theory. Unphysical features could be expected
to arise in this limit, because,
as we commented in the previous section, $|r| > 1$ would be forbidden by
the requirement of Osterwalder-Schrader positivity at the lattice level. However,
it can be checked that the meson propagators in this limit satisfy the
positivity condition. 

If we solve in this even-dimensional case for the critical hopping $\kappa_c^2$ which defines the critical
line as a function of $r$ we get, in particular, the result $\kappa_c^2(r=1)=1/4d$,
in agreement with the well-known $\kappa_c^2(r=1)=1/16$ for $d=4$ \cite{Kawam}-\cite{Aoki2}.

For odd space dimensions the previous analysis has to be modified. Now,  it is
not possible to satisfy
$|\tau|=d-1$ and $\varphi_i=0$ simultaneously, due to the requirement that
the number of
gamma-fermions must be even. Thus, the lightest state has a higher mass.
This state is a mixed vector-tensor state ($\gamma_i$ and $\gamma_0
\gamma_i$). The critical line is given by $|\Phi_2(\xi)- (d-1) \theta|=1$
and corresponds to a massless vector meson.
For $r > 1$ this massless  state is obtained for $\vec{\varphi}= \vec{0}$
and for $r < 1$ it corresponds to  one direction being $\varphi_i=\pi$.
There are no other massless particles except for $r \to \infty$. 

\vskip 1cm

After this general presentation, we will now restrict ourselves to the most
 interesting cases of 3 and 4 dimensions, where the continuum theory
is supersymmetric.  We will start by
considering $d=4$. The formulae for the propagators and masses  
follow from the general formulae obtained previously. The normalisation of
our operators in this case is given in Table 1. As mentioned previously, 
if we consider correlation functions  of ultralocal operators the only
non-zero ones for Majorana fermions are the scalar (S) and the
pseudoscalar-axial (PA) sectors. For ease of access to  the results we now 
collect the expressions  for  the  scalar mass $M_S$ and the lightest pseudoscalar
mass $M_P$:
\bea
\cosh(M_{S})&=&|\Phi_2(\xi) \pm 3| \no
\cosh{(M_{P})}&=&\theta\, \Xi - \sqrt{(\theta^2 -1)(\Xi^2-1) +
9 \epsilon^2} \,\,\, ,
\eea
where the $\pm$ in $\cosh(M_S)$ is for $|r|<1$ and $|r|>1$, respectively.
We recall the  expression for $\Phi_2$ and $\Xi$ in this dimension:
\bea
\Phi_2(x)&=&\frac{8x^2-14x+7}{2x(1-x)} \nonumber \\
\Xi&=&\Phi_2(\xi)-3r^2\epsilon \nonumber \, \,\,\, .
\eea
$\xi,~\epsilon$ and $\theta$ are, as usual, the ones in Eqs.~(\ref{xidef}, \ref{defvaria}).

The critical line where the pseudoscalar particle becomes massless is given
by 
$\kappa=\kappa_c(r)$, where the critical hopping parameter
$\kappa_c^{2}$ is given by:
\be
\kappa_c^{2}(r)=\frac{23r^2+9+3\sqrt{9r^4+46r^2+9}}{896r^4} \ \ .
\label{crithop}
\ee

Our formulae are equal for fields in the fundamental and the adjoint
representation of the group, so that up to a normalisation factor, they can
be directly compared with the results obtained previously on the literature
~\cite{Kawam}-\cite{Aoki2}. Once the appropriate normalisation of the fields
is taken into account, we find perfect agreement with the results obtained
by previous authors.

Now we discuss the $d=3$ case, for which no explicit results have
been given previously. The expression for the propagator does not
follow from our previous formulae. In this case the basis of
meson operators is given by the set of 2 by 2 matrices $\Sh_i=\{{\bf
1},~\sigma_i\}$, where $\sigma_i$  are the Pauli matrices.
To comply  with the conventional indexing of the Pauli matrices, we 
will adopt the names $(\varphi_1,\varphi_2,\varphi_3)$
for the lattice momentum coordinates, time being assigned by convention to
coordinate 1.
It is easy to see that here the propagator splits into two separate blocks:
the scalar and the vectorial sector, corresponding to the 
unit matrix and Pauli matrices respectively.
By inverting the matrix $\Theta_2(\xi)\mathbf{I} - \tAtwo(\varphi)$
we find the following results for the scalar 
$\widehat{G}_{SS}(\varphi)$ and 
vectorial $\widehat{G}_{VV}(\varphi)$ propagators:
\bea
\widehat{G}_{SS}(\varphi) &=& \frac{H(\xi)}{\Phi_2(\xi)-\sum_{i=1}^{3} 
\cos({\varphi_i})}\no
\widehat{G}_{VV}(\varphi)&=&\frac{H(\xi) \hat{M}^{V}}{ \tilde{\alpha}_1
\tilde{\alpha}_2 \tilde{\alpha}_3 + \sum_{i=1}^3 \tilde{\beta}_i^2
\tilde{\alpha}_i  } \ \ .
\eea
where the matrix elements of $\hat{M}^V$ on the $\sigma_i$ basis are given by
\bea
\hat{M}^{V}_{ii}&=&\tilde{\beta}_i^2+\frac{1}{2}\sum_{j,k=1}^3 
(\epsilon_{ijk})^2 \tilde{\alpha}_j \tilde{\alpha}_k \nonumber \\
\hat{M}^{V}_{i\ne j}&=& \tilde{\beta}_i \tilde{\beta}_j-\sum_{k=1}^3\epsilon_{i j k} 
\tilde{\alpha}_k \tilde{\beta}_k \label{prop3d}\,\, ,
\eea
$\epsilon_{i j k}$ is the completely antisymmetric tensor 
and the functions $\tilde{\alpha}_i$ $\tilde{\beta}_j$ are given by
\bea
\tilde{\alpha}_i&=&\Phi_2(\xi)-\theta \sum_{i=1}^3 \cos({\varphi_i}) +
2 \epsilon \cos({\varphi_i})
\no
\tilde{\beta}_i&=&2\epsilon~r \sin({\varphi_i})\ \ .
\eea
For  $d=3$, the functions $H$ and $\Phi_2$ adopt the form:
\bea
H(x)&=&-\eta~D_R~\left(\frac{5}{2}\right) \frac{4x-5}{x(x-1)(6x-5)} \nonumber \\
\Phi_2(x)&=&\frac{6x^2 -10x+5}{2x(1-x)} \,\,\, .
\eea

Let us now consider meson masses. The general formulae~(\ref{mscal})-(\ref{massmix}) are
valid for
this case. The restriction to gamma-fermion states which are eigenstates of
the  charge conjugation operator $\mathbf{C}$ simply eliminates the double
degeneracy of all levels. Taking into account the necessary evenness of the
number of gamma-fermions, we arrive at:
\bea
\cosh(M_{S})&=& \vline \,\Phi_2-\sigma \,\vline \, = \, \vline \,\Xi + \sigma\epsilon \,\vline\\
\cosh(M_{11})&=&\vline \,\Phi_2-\sigma\theta \,\vline \, = \, \vline\, \Xi -\sigma\epsilon \,\vline \\
\cosh(M_{mix}^{\pm})&=& \theta \, \Xi \pm \sqrt{(\Xi^2-1)(\theta^2-1)+ \epsilon^2 (4-\sigma^2)} \,\,\, ,
\label{mass3d}
\eea
where, as usual, $\sigma=\sum_{i=2}^3\vartheta_i$, and
$\vartheta_i \equiv \cos({\varphi_i})$ takes values $\{-1,1\}$.
$\Xi$ is defined below Eq.~(\ref{massmix}).
The analysis of the lightest meson follows the general presentation done 
previously for odd space-time dimension. 

Finally, we will consider  correlation functions of the 3-gluino  operators
given in Eq.~(\ref{threegops}).
We recall that depending on the choice of the invariant colour tensor ($f$
or $d$) the spin matrix $\cS_i$ can be chosen completely symmetric or
completely antisymmetric under the  exchange of the 3 spinorial indices.
The general expression for the propagator was given in Section 2
(Eq.~(\ref{tgluino})), and involves the inversion of the matrix
$\Theta_3(\xi)\mathbf{I}- \tilde{\mathbf{A}}_3$. Although performing this
inversion explicitly in general seems a fairly complicated problem,
we will be able to perform it inversion for the simplest case.
This corresponds to a completely antisymmetric  $\cS_i$ matrix. Actually,
for $d=4$ it gives the full antisymmetric subspace.
We proceed  by introducing the antisymmetric matrices $\VM_A$, where
$A$ is an spinorial index which can be looked at as the spin components of
a composite fermion (a spin $1/2$ fermion in 4 dimensions). 
The form of $\VM_A$ is given by: 
\be
(\VM_A)^{A_1 A_2 A_3} = C^{A_1 A_2} C^{A_3 A} -
C^{A_1 A_3} C^{A_2 A} + C^{A_2 A_3} C^{A_1 A} \,\,\, ,
\ee
The advantage of these matrices  is that the states  $\CTE^{-1} | \VM_A
\rangle$ are the basis of an invariant subspace under the action of
$\tilde{\mathbf{A}}_3$. This allows us to perform the appropriate
inversion in this subspace. If we label by $\widehat{G}^{({\cal A})}_{A A'}$
the momentum space propagator between the states $\cS_i=\VM_A$ and
$\cS_j=\VM_{A'}$, we can write:
\bea
\widehat{G}^{({\cal A})}_{A A'} &=& -R_3(\xi) \, (6-3
Tr(\mathbf{I}_{spin})) \times \nonumber \\ 
&& \left( C \lbrack (\Theta_3(\xi) - 2 \kappa^3 
r(r^2 -1) \tilde{\sigma})\mathbf{I} + 2 \imath \kappa^3(r^2 -1) \sum_{\mu}
\sin(\varphi_{\mu}) \gamma_{\mu} \rbrack^{-1} \right)_{A A'}\ \ ,       
\eea
where $Tr(\mathbf{I}_{spin})$ is the dimension of the spin space (equal to
$2^{d/2}$ for even $d$)  and $\tilde{\sigma}=\sum_{\mu} \cos(\varphi_{\mu})$. The inversion of the
matrix contained in the previous formula  can be done in the standard way for  a fermion
propagator: $(a+ b_{\mu} \gamma_{\mu})^{-1}= (a- b_{\mu} \gamma_{\mu})/(a^2 -
b_{\mu}^2)$.  

Setting as usual the space-like momenta to $0$ or $\pi$, we might extract
the mass of this fermion state from the position of the pole in the 
propagator. The result is:
\be
\cosh(M_{\frac{1}{2}})= \ \vline r \Xi_3 \pm \sqrt{ (\Xi_3)^2-\epsilon} \ \vline \ \ ,
\ee
where we recall that $\epsilon = 1/(r^2-1)$ and the symbol $\Xi_3$ stands 
for:
\be
\Xi_3 = \frac{\Theta_3(\xi) \epsilon^2}{2 \kappa^3} - r \epsilon
\sigma\ \
\ee
with  $\sigma=\sum_i \cos(\varphi_i^{(special)})$.
It can be shown that, although the mass formula  depends on the sign of
$\kappa$ and $r$, the spectrum does not. Without loss of generality, we can
set $\kappa > 0 $; then,
the lowest mass is obtained for $\vec{\varphi}=\vec{0}$ when $r>1$ or $-1<r<0$,
and for $\varphi_i=\pi~\forall i$ when $r<-1$ or $0<r<1$.

Obtaining the remaining masses of 3 gluino states analytically in general is a
complicated problem. Nevertheless, we do not need them for the physical conclusions
of this paper. Indeed as we saw in the case of mesons the only point in
parameter space were the scalar and the pseudoscalar became degenerate in
mass occurred for $\kappa \to 0$, $r \to \infty$ with $\kappa r =
\sqrt{\lambda}$ fixed. Extracting the masses and propagators for all
$p$-gluino states in this limit is fairly simple. As in the case of mesons all
spin states are degenerate, and the masses only depend on $\sigma$. 
By looking at the expressions~(\ref{thetap},\ref{atp}) for
the matrix to be inverted when considering a $p$-gluino operator, we immediately
see that:
\be
\Theta_p \mathbf{I}- \tilde{\mathbf{A}}_p \stackrel{r \to \infty}{\longrightarrow} (\Theta_p -2 \kappa^p r^p \tilde{\sigma})\mathbf{I} \,\,\, .
\ee
Now, the eigenvalues, whose zeroes gives the masses, are explicit. Their
value is given by:
\be
\label{massform}
\cosh(M_p)= \Phi\left(\left(\frac{(2d-1)(1-\xi)}{\xi}\right)^{p/2}\right)  -\sigma\ \ ,  
\ee
where we have  used the definition~(\ref{xidef}) of $\xi$ and introduced the function
\be
\Phi(x) = \frac{1}{2} \left( x + \frac{(2d-1)}{x} \right) \ \ .
\ee
 As $p$ increases $\left(\frac{(2d-1)(1-\xi)}{\xi}\right)^{p/2}$, the argument of the
function $\Phi$ in (\ref{massform}), increases. It is easy to see that
within the relevant range ($x \ge 1$) the function $\Phi(x)$ is monotonously
increasing. This allows us to prove the following inequality:
\be
M_p > M_q\ \ \ \mbox{for }\ \ p > q  
\ee
As a consequence for any value of $\xi$ in the physical range
$\lbrack 0,\frac{1}{2} \rbrack$ and $\varphi_i=0$, the mass of the 3-gluino states is higher than
the mass of the 2-gluino states. In particular at the critical point $\xi=\frac{1}{2}$,
the $3$-gluino states are massive.

It is also interesting to know whether there are points within the physical
region  (bounded by the 2-gluino critical lines) where  some 3-gluino states
become massless. For example, in Ref.~\cite{Minkowski} it is predicted that 
if chiral symmetry is broken spontaneously then so is supersymmetry, and a massless
goldstino particle appears. We have explicitly checked that this does not occur 
within our framework for  the 
3 and 4-dimensional cases. The 3-gluino states are always massive.

%% file: section4.tex
\section{Conclusions and future outlook}
In this section we will summarise our results and discuss their
implications. We have obtained  the spectrum of
 N=1 SUSY Yang-Mills on the lattice at  large number of colours $\NC$ and
strong coupling, by considering the hopping parameter expansion as a
sum over lattice paths (random walks).  We have resummed the expressions
in the hopping parameter in a certain region enclosing the origin, for
an arbitrary  value of the Wilson parameter $r$. We have worked at zeroth 
order  in $\beta$, the pure gauge coupling constant. However,  
Wilson's  action for the gauge part can be added either as  a trace in the 
fundamental representation or as a trace in the adjoint one, with the
corresponding couplings related to match the same naive continuum limit. 
Indeed, if we  choose the adjoint version of Wilson action, our results 
(propagators and masses) are valid to all orders in $\beta$ (probably 
only within some region encircling the origin). This can be proven  in the 
same way that one sees that the  quenched approximation is exact in our case.

We have given formulae for the propagators and masses of 2 and 3 gluino
states. The 2-gluino masses do coincide with the results for the meson
spectrum in ordinary lattice QCD at strong coupling \cite{KawSmit}-\cite{Aoki2},
and obtained by means of the effective potential method. Our
method is based on the random walk resummation 
technique~\cite{resum},~\cite{combi}.
This generalises the method of Kawamoto~\cite{Kawam} for $r \ne 1$.
Both methods have their relative advantages and disadvantages and occasionally
there have been some conflicting conclusions (See the U(1) problem
discussion in Refs.~\cite{Aoki2} and \cite{Froh}). The random
walk method does not rely on specific assumptions about the symmetries
of the saddle point solution. It is rather based on the resummation of
convergent series. Convergence is simple to see, since the number of lattice paths
of length $L$ grows like a power of $L$ and the matrices involved are bounded.
The radius of convergence is given by the distance to the nearest
singularity. In our case, we have two resummations involved. One on the pure
spike paths, which converges provided $\kappa^2 < \frac{1}{4 (2d -1) (r^2
-1)}$, and a second one whose border line in four dimensions, is given by
the critical line (Eq.~(\ref{crithop})) where the {\em pseudoscalar} becomes
massless. Furthermore, in our case, the replacement of the Pfaffian by the
square root of the determinant can be rigorously justified if $|\kappa| <
\frac{1}{2d(|r|+1)}$. This falls a bit too short compared to the other
limits. Finally, we want to stress that we have provided formulae for
arbitrary space-time dimensions $d$, which could be of interest for other 
researchers in the field.  

Apart from the technical interest  of our methods and results, we consider
that the main usefulness of our results, is that they provide   a guideline
and boundary formulae for groups investigating this model by Monte Carlo
simulations. Of course, our results are only valid for large $\NC$, but
experience teaches us that this is frequently a numerically good approximation.
There is one issue in which unfortunately our method could perhaps not help. It has been
predicted, that this model should have a first order phase transition in
$\kappa$~\cite{montvay2}. Our work  predicts the presence of a second order
phase transition associated to the vanishing of the pseudoscalar mass. It can 
be argued however, that a series expansion like ours can be seen as the
expansion around one of the effective potential vacua. Thus, the mentioned
second order transition could lie in the metastable phase. The point about
the order of the phase transition should  be settled by future Monte Carlo 
simulations.

Finally,  it is tempting to speculate about the relevance of  our results
in the spirit of supersymmetry breaking. For that purpose one is interested in critical
lines where a continuum limit can be defined. The states whose
lattice mass vanish at the critical line, are the states that survive
this continuum limit. If supersymmetry  is recovered at this continuum
limit one expects these states to form  a supermultiplet. The
analysis of Curci and Veneziano leads to a  multiplet in
which  in addition to the pseudoscalar particle, there is   a scalar one and a
spin $\frac{1}{2}$ fermion. These particles have equal positive continuum masses.
Since the contribution to the mass of the pseudoscalar comes through the
anomaly, which vanishes in our case, we should expect a massless multiplet.
Nevertheless, it is doubtful that the analysis of Curci and Veneziano
applies at strong coupling since it is based on the continuum SUSY Yang-Mills
lagrangian. By power counting, this model has a single relevant parameter,
the gluino mass, and hence fine tuning one of the bare couplings one could
find a line corresponding to vanishing mass and restored supersymmetry.
However, at strong coupling the gluonic degrees of freedom stay of the order
of the cut-off. Hence, the low energy lagrangian, if supersymmetric,  would rather
coincide with the Wess-Zumino model. This has many more relevant operators (the
different masses and couplings) and  demands tuning of more bare parameters
to approach it. In this respect the situation in 3 dimensions could be
interesting since the model would be interacting. In 4 dimensions we would
expect a free low energy lagrangian giving the physics of the continuum
limit.  With all these concerns in mind we did not want to loose the
opportunity to explore the $\kappa-r$ parameter space in search for
degenerate low energy multiplets. Actually, we concluded that the  only point
where several mesons become  massless is in  the limit $\kappa \to 0$,
$r \to \infty$ and $r\kappa=\frac{1}{2\sqrt{2d-1}}$ (i.e., $\xi=\frac{1}{2}$).
The masses at this point are the ones corresponding to a gauge-Higgs system:
an interesting model in its own right.  At  this point we
looked at the masses of the  $p$-gluino states with $p>2$. This includes
fermionic degrees of freedom (for $p$ odd). However, we showed that these
states remain massive (in cut-off units) at this critical point. So that the
main conclusion is that there is no point in the $\kappa-r$ plane giving a
possible candidate for a supersymmetric continuum limit.

We conclude the paper by mentioning a few possible improvements of our
results. First of all, the possibility of extending the calculations and
formulae to next to leading order in $1/\NC$ seems a feasible one. The most
important consequence of this extension could be in cases when the effects
are absent to leading order, like the effect of quenching, anomalies, etc.
Then one can try to include higher orders in $\beta_{fundamental}$, or
combined $1/\NC$ and $\beta_{adjoint}$. Then, of course, it would be 
very good to rederive the results of this paper with the  effective action 
method. This technique, as mentioned previously,   would allow the  
discovery of possible first order phase 
transitions. Finally, it should be commented that our methods and results 
could be used to study other supersymmetric models, such as SUSY QCD.

%% file: acknow.tex
\section*{Acknowledgements}
We would like to thank J.M.F. Labastida for some clarifications 
concerning Supersymmetric lagrangians. 
E.G. would also like to thank C. Di Cairano for useful discussions.
This work has been partially financially supported by the 
CICYT under grant AEN97-1678. 
E.G. also acknowledges the financial support of the TMR network
project ref. FMRX-CT96-0090.

%% file: apendiceA
\section*{Appendix}
In this appendix  we will present the terminology and main 
results on random walks that we will need in the text.  
Not to conflict with other  definitions, we will  refer to these 
random walks as  {\em lattice paths} and a precise definition 
will be provided.  We will work in arbitrary dimension $d$ and will employ
additional definitions given at the beginning of Section 2.
Proofs will not be given. For that we address the reader to  Ref.~\cite{combi}.

A lattice path of length $L$ is an element 
$\gamma \equiv (n,\vec{\alpha}) \in {\cal L} \times I^L$. The point $n \in  {\cal L}$ is
the {\bf origin} of the path,  $\vec{\alpha}$ is the path {\bf sequence}, and 
$m=n+\sum_{i=1}^{L}\GV{\alpha_i}$ is  the {\bf endpoint} of the path. We can label
the different path sets as follows.  
 Let ${\cal S}_L(n)$ be the space of all paths with origin $n$ and length $L$;
  ${\cal S}_L(n\rightarrow m)$ is the space of all paths with origin $n$,
endpoint $m$ and length $L$. Now we will introduce the notion of a {\bf spike}.

A path $\gamma \equiv (n,\alpha_1,\ldots ,\alpha_L)$ has {\bf spikes}
if there exist one integer  $1 \le i \le L$ such that
$\alpha_{i+1}=\bar{\alpha_i}$. In the converse case one says that the path 
has  {\bf no spikes}. The set of all paths without spikes
of length $L$ and  origin $n$  is labelled  $\bar{{\cal S}}_L(n)$ 
($\bar{{\cal S}}_L(n\rightarrow m)$ if the endpoint is fixed to $m$).
Now, we can associate to any path $\gamma$ a corresponding path $\bar{\gamma}$
called its {\bf reduced path}, by simply eliminating in an orderly manner 
all pairs $\alpha_{i+1}=\bar{\alpha_i}$ in its sequence. Notice that this
procedure preserves the origin and endpoint of the path. 

Our main strategy will be to convert the sum of paths into a sum over 
reduced paths. For that we would first need to compute the following 
function:
\be
\label{Fdef}
F(\bar{L},z)= \sum_{p=0}^{\infty}\, z^p\, N(\bar{L},p) \ \ ,
\ee
where $N(\bar{L},p)$ is the number of paths of length 
$\bar{L}+2p$ whose reduced path is a path (with no spikes) 
of length $\bar{L}$. This number does not depend on the path 
itself but only on its length. In Ref.~\cite{combi} it is shown that:
\bea
  \label{expF}
   F(\bar{L},z)=\frac{1}{(1-\frac{2d}{2d-1}\xi)}
   \frac{1}{(1-\xi)^{\bar{L}}}\\
\label{defxi}
  \xi=\frac{1-\sqrt{1-4(2d-1)z}}{2} \ \ .
\eea

In addition, we would need to be able to perform sums over the set of 
reduced paths  with origin in $n$ of a product of matrices. To be specific,
let  $\mathbf{A}_{\alpha}$ for $\alpha \in I$, be a collection of matrices 
satisfying  $\mathbf{A}_{\alpha} \mathbf{A}_{\bar{\alpha}}=\lambda
\mathbf{I}$, with $\mathbf{I}$ the unit matrix. Then we want to compute the
matrix:
\be
\label{defT}
{\cal T}(\mathbf{A}) = 1 + \sum_{\bar{L}=1}^{\infty} \ \ \sum_{(n,\vec{\alpha})\in
\bar{\cal S}_L(n)}
\mathbf{A}_{\alpha_1} \cdots  \mathbf{A}_{\alpha_{\bar{L}}} .
\ee
It can be shown~\cite{combi} that ${\cal T}(\mathbf{A})$ is given by the
formula:
\be
\label{resT}
{\cal T}(\mathbf{A}) =(1-\lambda)(1+(2d-1)\lambda - \tilde{\mathbf{A}})^{-1} \ \ ,
\ee
where $\tilde{\mathbf{A}}=\sum_{\alpha\in I} \mathbf{A}_{\alpha}$.

In the text we will need to evaluate an expression like Eq.~(\ref{defT})
with a slight variation. We would need to sum only over paths going from one 
lattice point $x$ to another one $y$. This modified situation can be reduced 
 to the case given before by the following procedure. Instead of considering 
 the matrix $\mathbf{A}_{\alpha}$ we will multiply it by a phase $e^{\imath
\varphi_{\alpha}}$, where $\varphi_{\bar{\mu}}=-\varphi_{\mu}$. Then,
diagrams that go from $x$ to $y$ have coefficients that go like 
$e^{\imath \varphi (x-y)}$. Thus, the required  expression can be obtained 
by projecting onto this term:
\be
{\cal T}_{x \rightarrow y}(\mathbf{A})= \prod_{\mu} \left( \int
\frac{d\varphi_{\mu}}{2 \pi}\right)\, e^{-\imath \varphi (x-y)}\,
(1-\lambda)(1+(2d-1)\lambda - \tilde{\mathbf{A}'}(\phi))^{-1} \ \ ,
\ee
with $\tilde{\mathbf{A}'}(\phi)=\sum_{\alpha\in I} e^{\imath
\varphi_{\alpha}}\, \mathbf{A}_{\alpha}$.

%% file: tabla1.tex
\begin{tabular}{|c|c|c|c|}
\hline 
$i$ label & ${\cal{S}}_i$ & ${\cal{C}}_i$  & \# d.f. \\   

\hline 
 
$S$ & $\frac{1}{2} C \mathbf{I}$ & $\mathbf{I}$ & $1$ \\
$V(\rho)$ & $\frac{1}{2} C \gamma_{\rho}$ & $\mathbf{I}$ & $4$ \\
$T(\rho \sigma)$ & $\frac{-\imath}{4\sqrt{2}} C \lbrack \gamma_{\rho},\gamma_{\sigma} \rbrack$ & $\mathbf{I}$ & $6$ \\
$A(\rho)$ & $\frac{\imath}{2} C \gamma_{\rho}\gamma_5$ & $\mathbf{I}$ & $4$ \\
$P$ & $\frac{1}{2} C \gamma_5$ & $\mathbf{I}$ & $1$ \\

\hline

\end{tabular}